      [1999/12/01 v1.4c Il Nuovo Cimento]
\documentclass{cimento}

\usepackage{graphicx}  % got figures? uncomment this
\title{Single and Double 
Top Quark Production at the Tevatron}%
\author{Daniel Wicke\from{ins:Mz}
}
\instlist{%
\inst{ins:Mz} Johannes Gutenberg Universit{\"a}t Mainz \-
  Staudingerweg 7,  55099 Mainz, Germany
\\  for the CDF and D0 collaborations
}
\PACSes{
\PACSit{14.65.Ha}{Top quarks}
\PACSit{14.70.Pw}{Other gauge bosons}
}
\newlength{\ziffer}

\newlength{\vorzeichen}

\newcommand{\TeV}{\,\mbox{Te\kern-0.2exV}}
\newcommand{\GeV}{\,\mbox{Ge\kern-0.2exV}}
\newcommand{\mGeV}{\,\mathrm{Ge\kern-0.2exV}}
\newcommand{\MeV}{\,\mbox{Me\kern-0.2exV}}
\newcommand{\keV}{\,\mbox{ke\kern-0.2exV}}
\newcommand{\eV}{\,\mbox{e\kern-0.2exV}}

\newcommand{\ifb}{\,\mbox{fb}^{-1}}
\newcommand{\pb}{\,\mbox{pb}}

\newcommand{\bea}{\pagebreak[3]\begin{samepage}\begin{eqnarray}}
\newcommand{\eea}{\end{eqnarray}\end{samepage}\pagebreak[3]}
\newcommand{\beq}{\begin{equation}}
\newcommand{\eeq}{\end{equation}}

\newcommand{\met}{\slash\!\!\!\!{E_T}}

\newcommand{\abb}{Fig.~\ref}
\newcommand{\fig}{\abb}

 \newlength{\howlong}

\begin{document}

\maketitle

\begin{abstract}
The CDF and D0 experiments have measured single and double top quark
production in $p\bar p$ collisions at the Tevatron 
at a centre-of-mass energy of $1.96\TeV$.
The applied methods are used to constrain properties of the top quark and to
search for new physics. 
Several methods of signal to background separation and of the
estimation of the background contributions are discussed. Experimental results
using an integraged luminosity up to $5.3\ifb$ are presented.
\end{abstract}
\section{Introduction}
Since the top quark was discovered by CDF and D0 at the Tevatron in 
1995~\cite{Abe:1995hr,Abachi:1995iq} the
number of top events available for experimental studies has been
increased by more than an order of magnitude. Tevatron delivered
a luminosity of more than $7\ifb$ and up to
$5.3\ifb$ have been used 
for top quark analyses in CDF and  D0.

In the Standard Model (SM) top quarks can be produced  through the strong or through the
weak interaction. The strong interaction creates top quarks in pairs.
The process is expected to be dominated by quark
anti-quark annihilation with a contribution of only $15\%$ from the gluon
fusion processes. The cross-section for this process is around $7\pb$. In the weak interaction top quarks
can be produced singly. At the Tevatron the two mechanisms known as
$s$-channel and the $t$-channel contribute to single top production 
with the ratio $1:2$ to a total of about $3.5\pb$.

Immediately after production top quarks decay to a $W$ boson and a $b$ quark
with a branching fraction of nearly $100\%$. The
decay channels of top quark pairs are thus fully specified through the $W$
boson decay modes. For the top pair production dileptonic decays including 
electrons and muons allow for the
highest purity, but  suffer from the low branching fraction of about 5\%.
The semileptonic decays are considered as the golden channel due to a sizable
branching fraction combined with the possibility to reach 
a reasonable signal to background ratio. The all-hadronic decay channel has the
largest cross-section, but due to the absence of leptons it suffers from a
huge background due to multijet production. Channels including $\tau$ leptons
are kept separately due to the difficulties in their identification.
For single top production the events are categorised by production and decay
channel. So far only leptonic decays were studied.

In the following first some new results on the top quark pair production
cross-section are described, followed by a selection of related and derived
results.
Then the observation of single top production and some related results are discussed.
A discussion of measurements of the top quark mass and further top quark
properties can be found in~\cite{GabrieleCompostella}.

\section{Top Pair Production Cross-Section}

The total cross-section of top pair production has been computed in perturbation theory using various
approximations~\cite{Kidonakis:2003qe,Cacciari:2008zb,Moch:2008qy,Moch:2008ai,Kidonakis:2008mu}.  
For a top quark pole mass of $172.5\GeV$ Moch
and Uwer~\cite{Moch:2008ai} find $\sigma_{t\bar t}=7.46^{+0.48}_{-0.67} \pb$,
based on the CTEQ6.6~\cite{Pumplin:2002vw} PDF. 
Experimentally it is important to measure this value in various decay
channels. In addition some measurements are done requiring identified $b$ jets
while others  avoid $b$ jet identification and rely on topological selections.
Most analysis use sideband data to evaluate the normalisation of the important
background contributions.  In the lepton plus jets channel the precision is
already dominated by systematic uncertainties. 
In this channel production of $W$-bosons with additional jets yields the
dominating background. Due to the difficulties in computing absolute
cross-sections for this process at high accuracy it is important to take the
corresponding background estimate from data. 

A sizable contribution of the systematic uncertainties of these measurements 
also stems from the luminosity determination. To overcome this limitation CDF has
measured the ratio of top quark pair production to the $Z$ boson production
cross-sections~\cite{CdfNote9616,CdfNote9474}. In $4.6$ and $4.3\ifb$ of data
CDF finds $\sigma_{Z\rightarrow\ell\ell} / \sigma_{t\bar t}=35.7$ and 
$\sigma_{Z\rightarrow\ell\ell} / \sigma_{t\bar t}=33.0$ for the analysis
using $b$ jet identification and the topological analysis, respectively. 
In the analysis using identified $b$ jets, the background is normalised from
data without identification requirement. In the topological analysis this
normalisation is obtained from fits to the topological likelihood discriminant.
The cross-section ratios are converted to top quark pair production
cross-sections using the 
theoretical prediction for $Z$ boson production. The theoretical uncertainty
induced by this step is much smaller than the luminosity uncertainties and
thus yield results with an uncertainty comparable to the uncertainty on the
prediction for top quark pair production:
\bea
\quad\sigma_{t\bar t} &=&7.14\pm0.35_\mathrm{(stat)}
\pm0.58_\mathrm{(syst)}
\pm0.14_\mathrm{(theory)}\pb\quad\mbox{using $b$ jet identification,}
\nonumber \\
\quad\sigma_{t\bar t} &=& 7.63\pm0.37_\mathrm{(stat)}
\pm0.35_\mathrm{(syst)}
\pm0.15_\mathrm{(theory)}\pb\quad\mbox{using topological selection.}
\eea

D0 has recently published an analysis of the full hadronic channel. The
analysis requires 6 jets two of which need to be identified as $b$ jets. In this
channel the background is dominated by multijet production from gluons and
quarks other than the top quark. It is modeled from data with 4- and 5-jets by
adding jets taken from 6 jet events. Only jets with lowest (and second lowest)
$p_T$ in the event are taken from 6 jet event. They must remain the lowest (or
second lowest) jet in the newly constructed event. 
This method of event constructed has been validated by adding one jet to 4-jet
events and compare them to normal 5-jet events. 

The final cross-section is obtained by fitting a likelihood discriminant as
observed in data to the prediction for top quark pairs  from simulation and the
background model described above. In $1.0\ifb$ D0 finds~\cite{Abazov:2009ss}:
\beq
\sigma_{t\bar t}=6.9\pm 1.3 (\mathrm{stat})\pm 1.4
(\mathrm{syst}) \pm 0.4 (\mathrm{lumi})\pb~\mbox{.}
\eeq
\begin{figure}
  \centering
\includegraphics[width=0.35\textwidth]{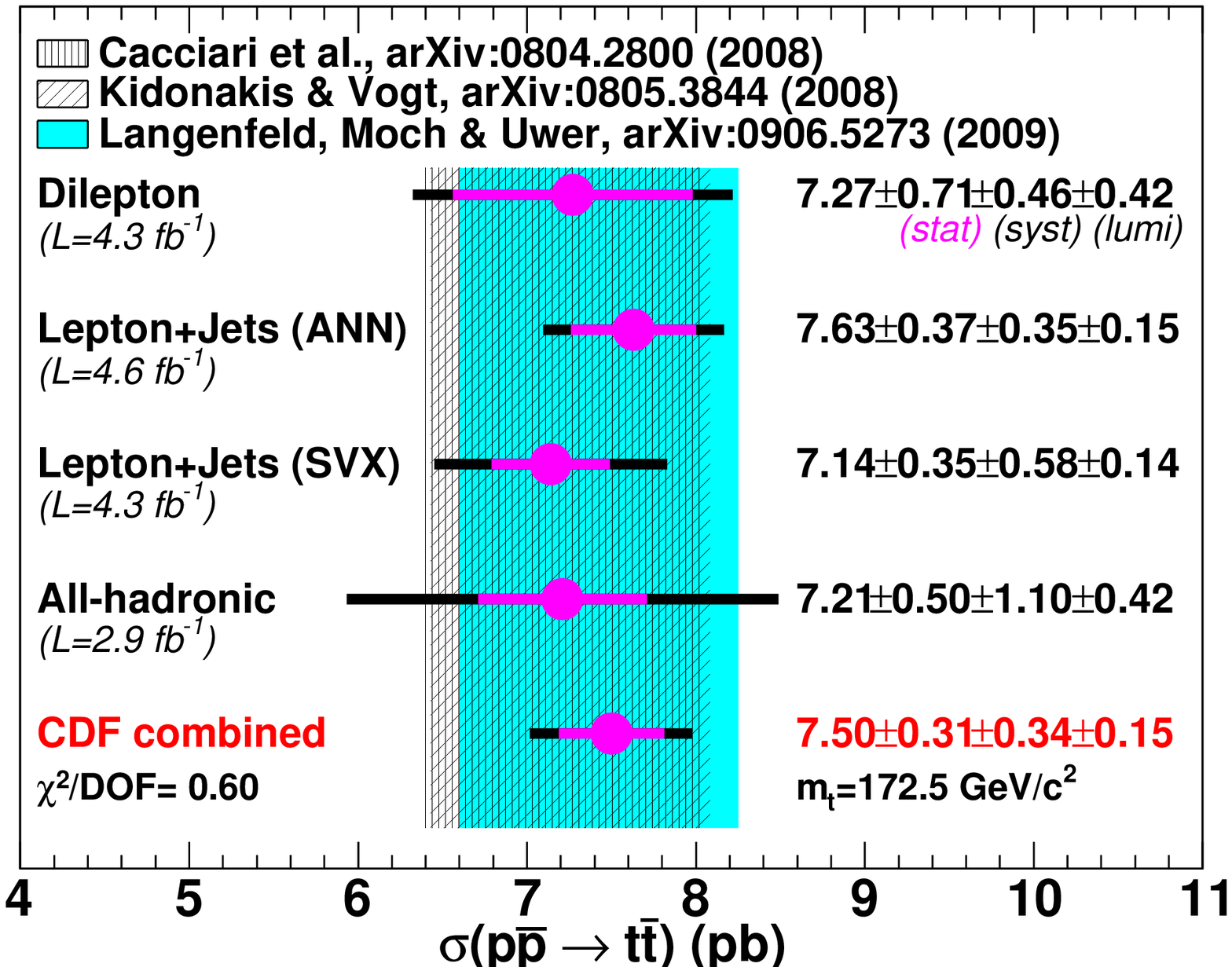}
\includegraphics[width=0.35\textwidth,clip,trim=0mm 4mm 0mm 0mm]{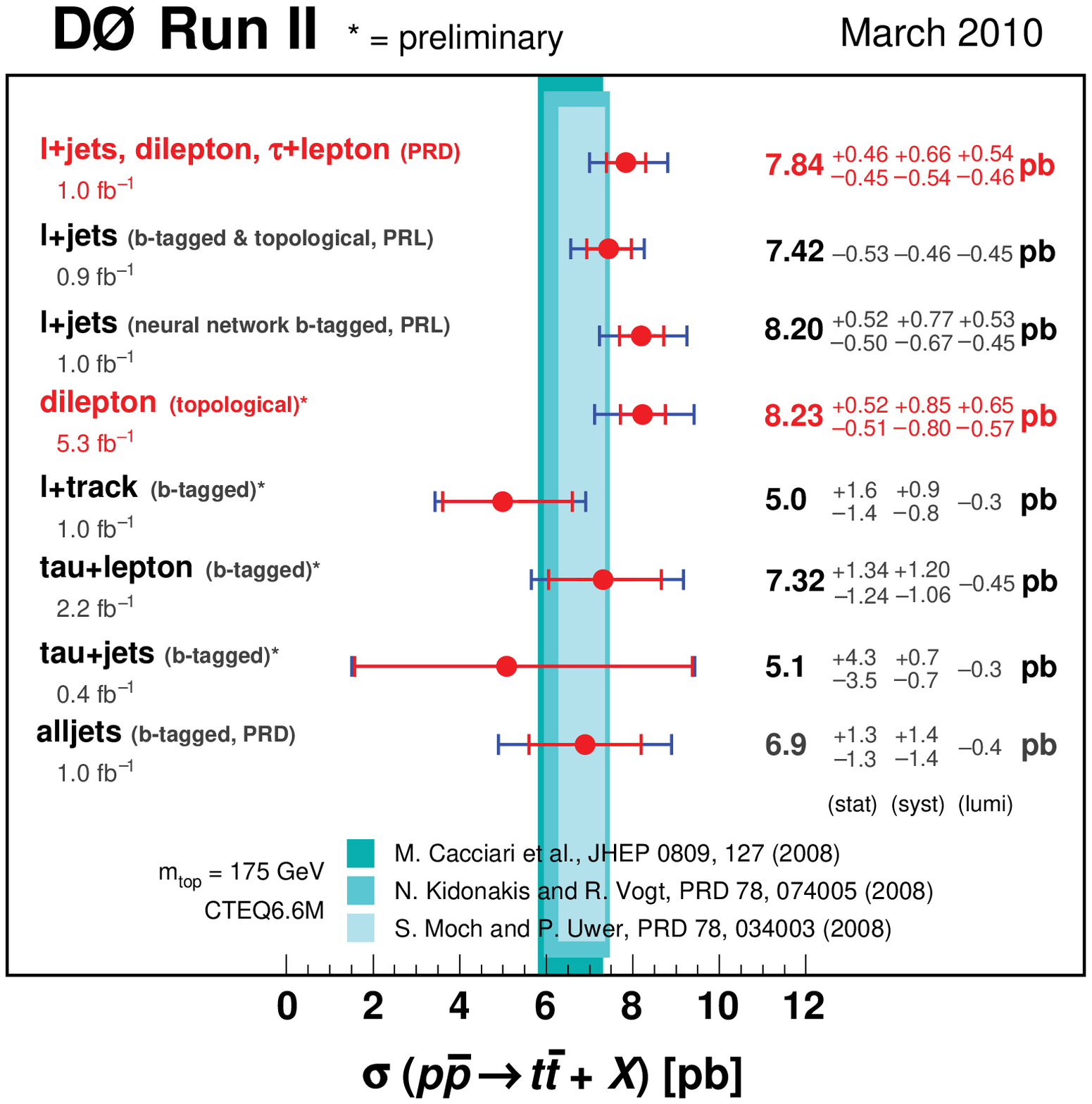}    
  \caption{Top quark cross-section measured in various 
    channels by CDF (left) and D0 (right).}
  \label{fig:xsec-summary}
\end{figure}%

Figure~\ref{fig:xsec-summary} summarises of the cross-sections for top quark
pair production  as measured by CDF and D0 in various channels including the
D0 dilepton result that was updated to $5.3\ifb$ since the conference~\cite{d0note6038conf}. All
measurements agree well with the theory predictions shown as vertical
bands. 

\subsection{Top quark mass from cross-section}
The theoretical predictions and (through the selection efficiency) also the
experimental results depend on the assumed top quark mass. This can be used to
determine the top quark mass in the well defined pole mass scheme. 
Comparing the experimental results from three channel ($\ell+$jet, the dilepton and the
$\tau+$other lepton) to  the prediction from Moch and Uwer~\cite{Moch:2008ai} D0 obtains
\beq
m_t^\mathrm{Pole}=169.1^{+5.9}_{-5.2} \GeV
\eeq
This results has larger experimental uncertainties but 
is consistent with direct mass determinations for which the mass
scheme, however, is not well defined. 

\subsection{Search for resonant top quark pair production}
In the SM no resonant top quark production is expected. However, unknown heavy resonances decaying to top
pairs may add a resonant part to the SM production mechanism. 
Such resonant production occurs in several models of new physics: 
e.\,g.~for massive $Z$-like bosons in
extended gauge theories \cite{Leike:1998wr}, for Kaluza-Klein states of the gluon
or $Z$ boson~\cite{Lillie:2007yh,Rizzo:1999en},  for axigluons
\cite{Sehgal:1987wi}, in Topcolor \cite{Hill:1993hs,Harris:1999ya}.  
D\O\ investigated the invariant mass distribution of top pairs in 
up to $3.6\ifb$ of $\ell+$jets
events~\cite{Abazov:2008ny,D0Note5600conf,D0Note5882conf}. 
Signal simulation is created for various resonance masses between $350$
and $1000\GeV$. The width of the resonances was chosen to be $1.2\%$
of their mass, which is much smaller than the detector resolution.
The top pair invariant mass, $M_{t\bar t}$, is reconstructed directly from the
reconstructed physics objects. A constrained kinematic fit
is not applied. Instead the momentum of the
neutrino is reconstructed from the transverse missing energy, $\met$,
which is identified with the transverse momentum of the neutrino. The
$z$-component is obtained  by solving
$M_W^2=(p_\ell+p_\nu)^2$ , where $p_\ell$ and $p_\nu$ are the four-momenta of
the lepton and the neutrino, respectively.

As the data agrees with the SM expectations, limits on the possible
contribution of resonant production $\sigma_{X}{\cal B}(X\rightarrow t\bar t)$
are set. The benchmark model  of Topcolor assisted Technicolor can be excluded
for $Z'$ masses of $M_{Z'}<820\GeV$.
A CDF study of $2.8\ifb$ in the all hadronic channel 
excludes $M_{Z'}<805\GeV$ in this model~\cite{CdfNote9844}.

\subsection{Unfolded differential cross-sections}
Besides the total cross-section in different channels, differential
cross-sections can be used to validate our understanding of top quark pair
production.

In CDF has recently published a measurement of the unfolded differential
cross-section with respect to the invariant top quark pair
mass, $\mathrm{d}\sigma_{t\bar t}/\mathrm{d}M_{t\bar t}$~\cite{Aaltonen:2009iz}.
Lepton plus four or more jet events are selected with at least one identified
$b$ jet using $2.7\ifb$ of CDF data. The invariant mass $M_{t\bar t}$ is
computed from the four leading jets, the lepton and the missing transverse
energy. The neutrino $z$ momentum is set to zero.
The expected background is subtracted from the observed distribution, then 
distortions are unfolded using the
singular value decomposition %~\cite{Hocker:1995kb} 
of the response matrix 
obtained from simulations. The final result is shown in \fig{fig:xsecmtt}~(left).
The consistency with the SM expectation 
is tested using Anderson-Darling statistics. %~\cite{AndersonDarling:1952}
The observed result
has a probability of 0.28 to occur if the SM is correct, showing good agreement of with the SM.
\begin{figure}[b]
\centering
\includegraphics[height=13.5em,clip,trim=0mm 0mm 0mm 3mm]{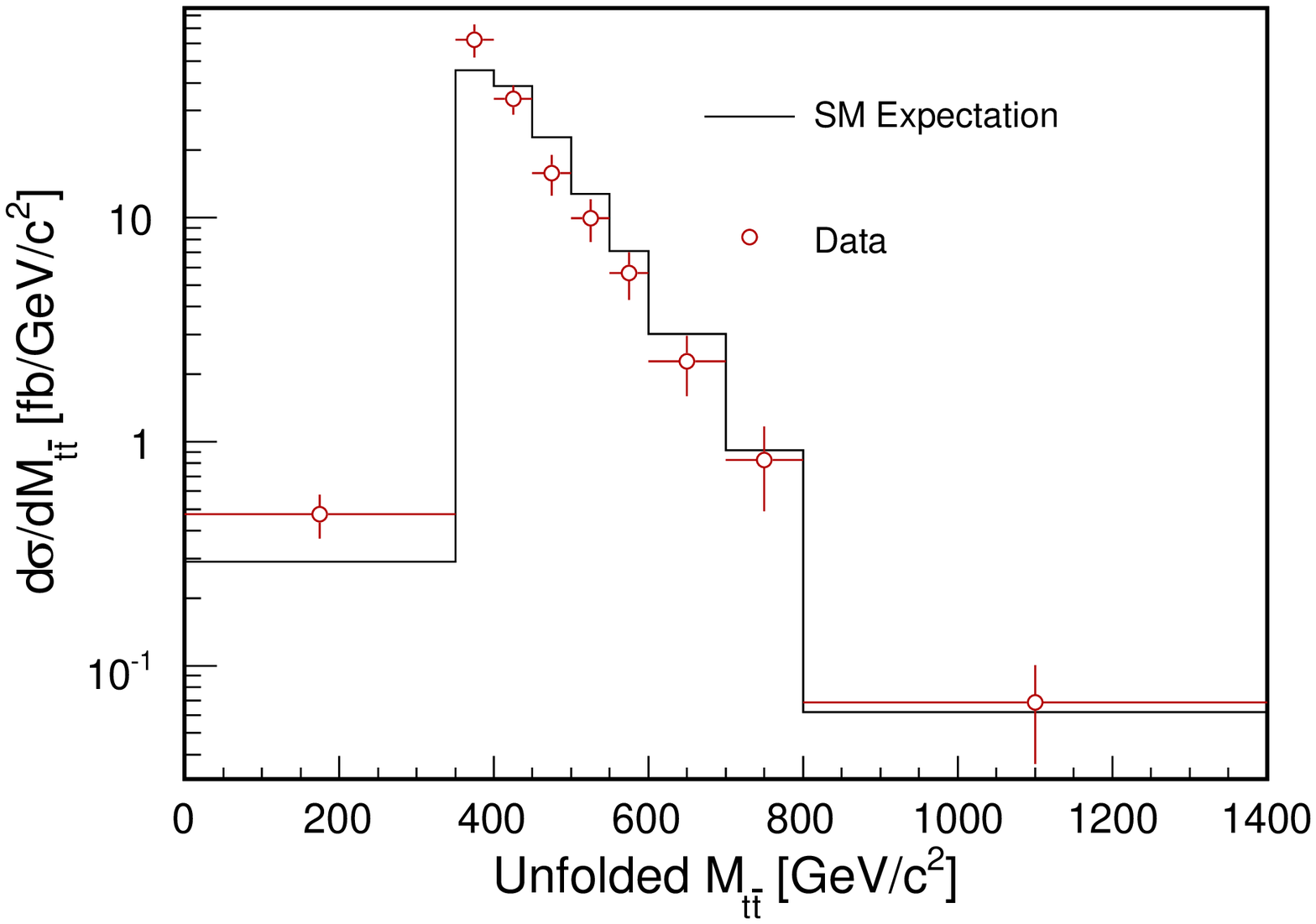}
\includegraphics[height=13.5em,clip,trim=0mm 0mm 0mm 7mm]{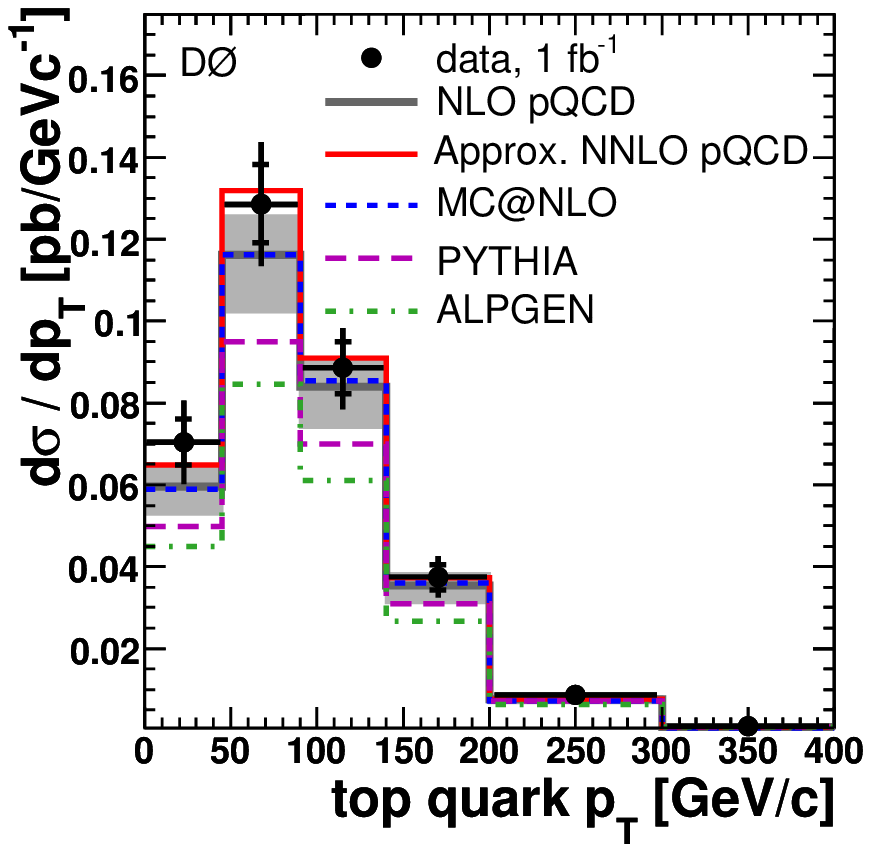}
 \caption{Differential cross-section of top quark pair production. Left as
   function of the invariant top quark pair mass measured by CDF~\cite{Aaltonen:2009iz} and right as function
   of the top quark transverse momentum by D0~\cite{Abazov:2010js}.}
 \label{fig:xsecmtt}\label{fig:xsecpt}
\end{figure}

D0 has determined the  unfolded differential
cross-section with respect to the top quark transverse momentum, 
$\mathrm{d}\sigma_{t\bar t}/\mathrm{d}p_T^t$ using $1.0\ifb$~\cite{Abazov:2010js}. In lepton plus jet
events including at least one identified $b$-jets the top quark transverse
momentum is reconstructed using a kinematic fit. The fit utilises the
measured momenta of the four leading jets, the charged lepton and the
missing transverse energy to determine the momenta of the top quark decay
products (four quarks, a charged lepton and a neutrino). Constraints on the
$W$-boson mass and on the equality of the top and anti-top quark masses are applied.
The expected background contributions are subtracted from the measured
distribution. Then regularised unfolding is used to determine the final
$\mathrm{d}\sigma_{t\bar t}/\mathrm{d}p_T^t$ shown in \fig{fig:xsecpt}~(right)~\cite{Abazov:2010js}. 
The result is compared to prediction of perturbative QCD (in approx.\ NNLO) and
various event generators. Perturbative QCD and MC@NLO show the best
agreement, but Pythia and Alpgen reproduce the observed shape at high $p_T$.

\section{Single Top Quark Production}
The cross-section for single top quark production is only half that of top
quark pair production. The same backgrounds as in top quark pair analyses
contribute and top quark pair production itself becomes a background. 
Moreover, in single top quark events have a signature containing
fewer jets than for top quark pairs. The signature selection requires an isolated lepton,
missing transverse energy and two to four jets, at least one of which must be
identified as $b$-jet. After this selection the signal to background ratio is
at best 1:10. Multivariate techniques are required to further separate
single top quark events from the backgrounds.

Both experiments employ multiple such methods, including boosted decision
trees, various neural network methods, matrix element and likelihood techniques.
The different multivariate methods are sensitive to different single top quark
events. Thus a combination of the different analyses improves the significance
of the result. The $5\sigma$ observations  that were reported in 2009 at this
conference have been published~\cite{Abazov:2009ii,Aaltonen:2009jj} and
combined cross-section~\cite{Group:2009qk} of
\beq
\sigma_t=2.76^{+0.58}_{-0.47}\pb
\eeq
is in good agreement with the SM expectations~\cite{Harris:2002md,Kidonakis:2006bu}.
\begin{figure}[b]
  \centering
  \includegraphics[width=0.6\textwidth]{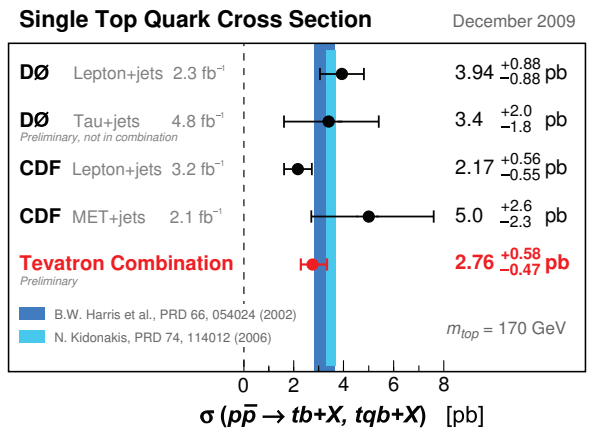}
  \caption{Single top quark cross-section measured by CDF and D0.}
  \label{fig:singletopxsec}
\end{figure}

In addition to the to the results obtained in the channels involving electrons
or muons (marked as ``Lepton+jets''), \fig{fig:singletopxsec} contains two
more recent results. One by CDF~\cite{Aaltonen:2010fs} with $2.1\ifb$  omits the explicit requirement for an
isolated electron or muon. This picks up events failing the lepton
requirements and events with taus in the final state. No explicit tau
reconstruction was done here. The second additional analysis done by D0~\cite{Abazov:2009nu} uses
boosted decision trees to explicitly recontruct hadronic tau decays. 
This reconstruction is trained individually  for three  tau decay modes that are
classified as $\tau\rightarrow \pi^\pm+\nu_\tau$, $\tau\rightarrow
\rho^\pm+\nu_\tau$ and $\tau\rightarrow 3\pi^\pm+\nu_\tau(+\pi^0)$. Signal
efficiencies between 59\% and 76\% are achieved at a background rejection rate
of 98\%. In $4.8\ifb$  D0 determines the single top cross-section in the
$\tau+$jets channel as $\sigma_t=3.4^{+2.0}_{-1.8}\pb$.

% 2 lines possible here

\subsection{Determination of $V_{tb}$}
The single top quark prodcution in the SM is directly proportional to the
CKM-matrix element $|V_{tb}|^2$. Thus the results presented above can be easily
interpreted as a measurement of  $|V_{tb}|$. For the combined result CDF and
D0 obtain $|V_{tb}|=0.88\pm0.07$. Constraining the value to the
allowed range between 0 and 1 yields a lower limit of $|V_{tb}|>0.77$ at 95\%
confidence level~\cite{Group:2009qk}. This limit is valid independent of the number of
generations, it however assumes $|V_{tb}|^2\gg|V_{td}|^2+|V_{ts}|^2$.
This assumption is supported by measurements in top quark pairs, see
e.g.~\cite{Abazov:2008yn}.

\subsection{Separation of $s$- and $t$-channel}
As explained in the introduction single top quark production at the Tevatron
actually consists of two separate processes, the  $s$- and the $t$-channel. 
The results presented so far consider the sum of the two
channels. CDF and D0 have also determined the two contributions separately~\cite{Abazov:2009pa,Aaltonen:2010jr}. 
The two dimensional results are shown in \fig{fig:singletopt}. The
CDF results shows a deviation to the SM expectation of a little more than two
standard deviations, while D0 result agrees very well with the SM. The
\fig{fig:singletopt}~(right) shows the D0 results for a discriminant that was
optimised to determine the $t$-channel cross-section. For this individual
channel D0 finds in $2.3\ifb$
\beq
\sigma_t^{t-\mathrm{channel}}=3.14^{+0.94}_{+0.80}\pb
\eeq
with a significance of $4.8$ standard deviation.
\begin{figure}[h]
  \centering
\includegraphics[width=0.4\textwidth]{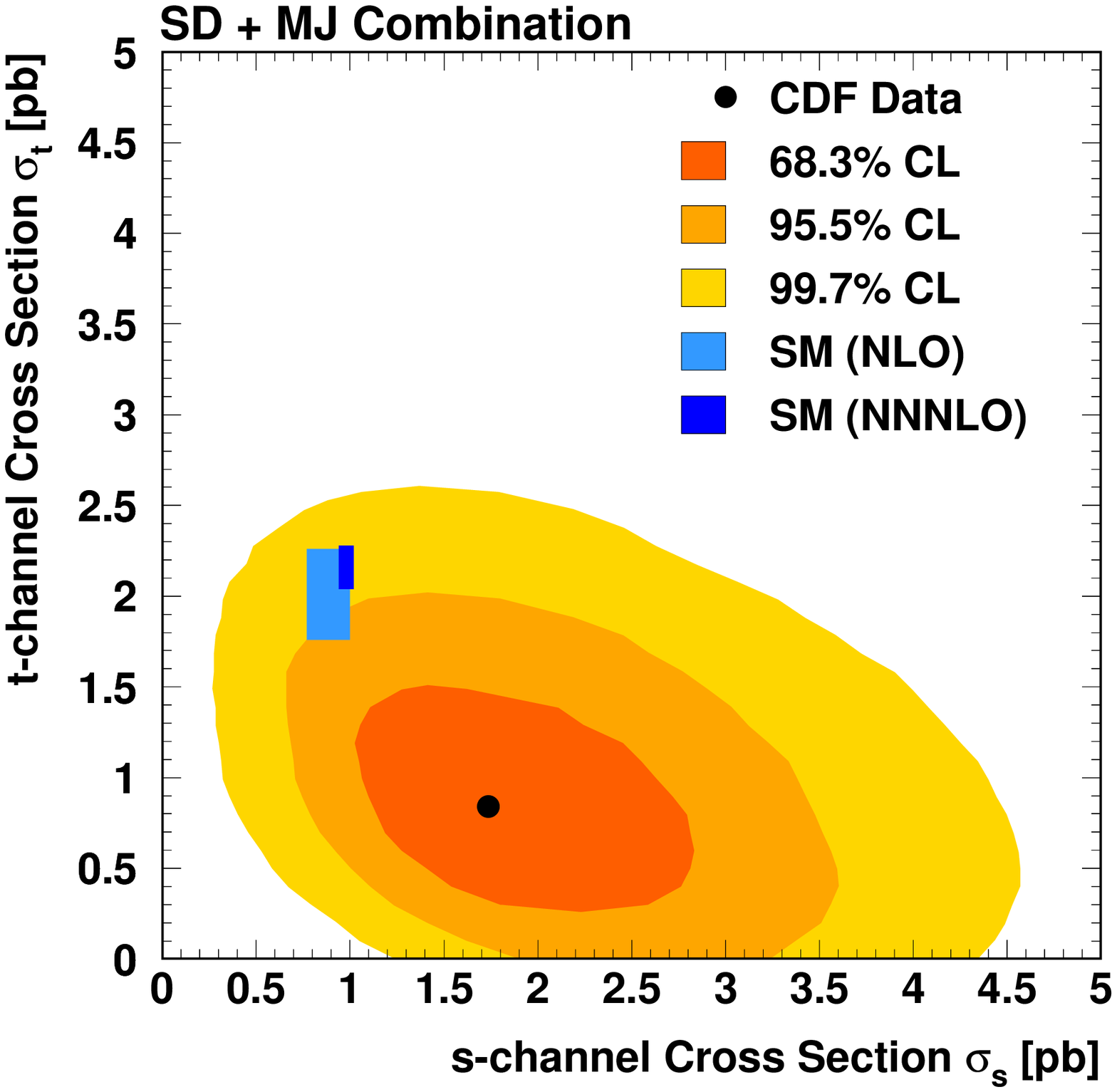}
%\hfill\includegraphics[width=8cm]{figs/T09IF1b.eps}
\includegraphics[width=0.4\textwidth]{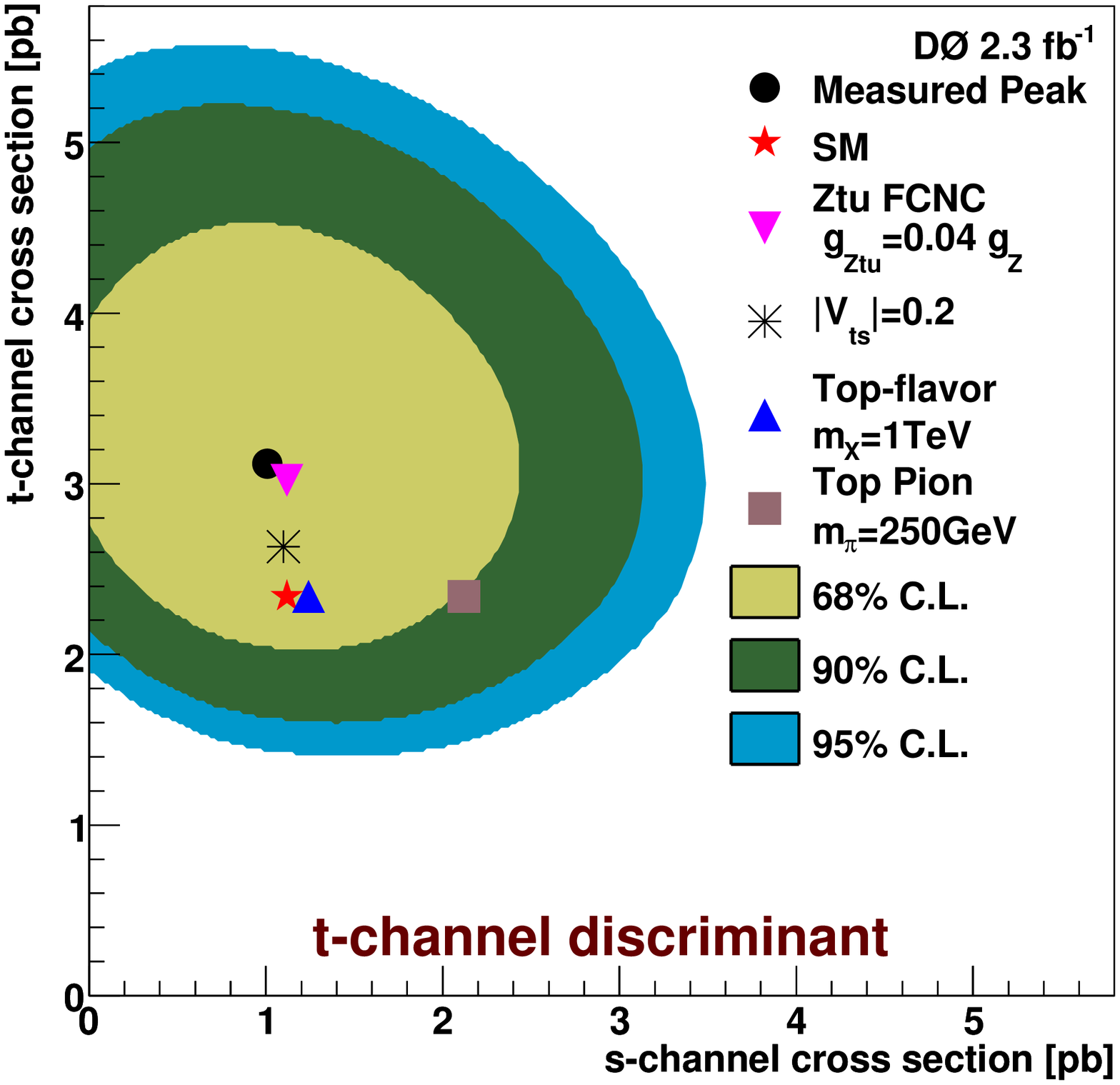}
  \caption{Simultaneaous fit of $s$ and $t$-channel contribution to the CDF
    (left) and D0 (right) data~\cite{Abazov:2009pa,Aaltonen:2010jr}. 
    Results are compared to the SM expecations and
    some selected alternative models~\cite{Tait:2000sh}.}
  \label{fig:singletopt}
\end{figure}

\subsection{Polarisation of the Top Quark}
In the presence of non-SM contributions to the top quark production~\cite{Tait:2000sh}, the
polarisation of the top quarks may be modified with respect to the SM
expectations. CDF considered a contribution of a production through
right-handed couplings, keeping the SM left-handed decay. Such a process could
be implemented through a heavy right handed $W^\prime$-boson. 
CDF trained their likelihood discriminant separately
for the the right-handed excotic and left-handed SM  case. With this the corresponding
two cross-sections $\sigma_R$ and $\sigma_L$ are measured and combined to a
polarsation
${\cal P}={(\sigma_R-\sigma_L)}/{(\sigma_R+\sigma_L)}$. In $3.2\ifb$ CDF obtains
${\cal P}=1.0^{+1.5}_{-0}$~\cite{CdfNote9920} in agreement with the pure SM
production.

\section{Conclusions}
The increasing Tevatron luminosity allows to measure the top quark
cross-section and properties with improved precision. 
The integrated 
top quark pair production cross-section is measured  in various decay channels
and used to obtain the top quark pole mass. 
Measurements of the differential cross-section as
function of $p_T$ or $M_{t\bar t}$ complement the verification of our
understanding of top quark pair production and are used to search for
deviations from the SM. Since the observation of single top quark production
at the previous La Thuile conference, new selection channels were added to the
studies and $s$- and $t$-channel contributions were measured separately.
In addition polarisation studies were studies in these events.

This note only describes
a small fraction of all measurements. The Tevatron experiments measure the
full spectrum of top quark properties to check the production, the decay and
inherent properties of the top quark against the SM expectation. So far no
evidence for new physics has been found.

\bibliographystyle{varenna-DW}
\begin{flushleft}
\bibliography{Tevatron,Habilitation,Dzero,CDF,local}
\end{flushleft}
\end{document}
\endinput
%%
%% End of file `cimsmple.tex'.